\documentclass[a4paper,11pt]{article}
\pdfoutput=1 
\usepackage{jcappub} 
\usepackage{lineno}
\usepackage{caption}
\usepackage[dvipsnames]{xcolor}
\usepackage{aasmacros}              
\usepackage{comment}
\usepackage{array}
\usepackage{subcaption}
\usepackage{multirow}
\usepackage{graphicx}
\graphicspath{{graphics/}}
\usepackage{amsmath}
\usepackage{amssymb}
\usepackage{rotating} 
\usepackage{xspace}
\usepackage{xcolor}
\usepackage{booktabs}
\usepackage[normalem]{ulem}
\usepackage{adjustbox}
\usepackage{lipsum}  
\newcommand{\Msun}{{\rm M}_\odot}

\title{Significant impact of Galactic dark matter particles on annihilation signals from Sagittarius analogues }

\author[a]{Evan Vienneau,}
\author[b]{Addy J. Evans,}
\author[b,a]{Odelia V. Hartl,}
\author[a, c]{Nassim Bozorgnia,}
\author[b]{Louis E. Strigari,}
\author[d]{Alexander H.~Riley,}
\author[e]{Nora Shipp}

\affiliation[a]{Department of Physics, University of Alberta,
CCIS 4-181, Edmonton, Alberta T6G 2E1, Canada}
\affiliation[b]{Department of Physics and Astronomy,
Mitchell Institute for Fundamental Physics and Astronomy,
Texas A$\&$M University, College Station, TX 77843, USA}
\affiliation[c]{Theoretical Physics Institute, University of Alberta,
CCIS 4-181, Edmonton, Alberta T6G 2E1, Canada}
\affiliation[d]{Institute for Computational Cosmology, Department of Physics, Durham University, Durham DH1 3LE, UK}
\affiliation[e]{McWilliams Center for Cosmology, Department of Physics, Carnegie Mellon University, 5000 Forbes Ave, Pittsburgh, PA 15213, USA}

\abstract{We examine the gamma-ray signal from dark matter (DM) annihilation from analogues of the Sagittarius (Sgr) dwarf spheroidal galaxy in the Auriga cosmological simulations. For velocity-dependent annihilation cross sections, we compute emissions from simulated Sgr subhalos and from the Milky Way (MW) foreground. In addition to the annihilation signals from DM particles bound to Sgr, we consider for the first time the annihilation of DM particles bound to the MW that overlap spatially with Sgr. For p-wave models this contribution can enhance the signal by over an order of magnitude, while for d-wave models the enhancement can be over three orders of magnitude. For Sommerfeld and s-wave models, the corresponding emission does  not significantly change. For the Sommerfeld model, the Sgr source can be visible above the MW foreground emission, while for s, p and d-wave models, the signal towards Sgr is most likely dominated by foreground MW emission. We interpret our results within the context of the observed gamma-ray emission from Sgr. We find that, given the background emission estimated from this region, the templates from simulations likely have spatial morphology that is too extended to explain the point-like emission that is observed.  
} 

\begin{document}
\maketitle
\flushbottom

\section{Introduction}
\label{sec:intro}

\par Dwarf spheroidal galaxies (dSphs) provide unique targets for dark matter (DM) annihilation~\citep{Tyler:2002ux,Evans:2003sc,Strigari:2006rd}. Nearly all dSphs are DM dominated and contain no intrinsic astrophysical sources of gamma-ray emission~\cite{Conrad:2015bsa,Strigari:2018utn}. The lack of significant signal from the most well-studied dSphs has established the most stringent limits on the annihilation cross section~\cite{Fermi-LAT:2011vow,Geringer-Sameth:2014qqa,Fermi-LAT:2015att,Fermi-LAT:2016uux,McDaniel:2023bju}. 

\par Of the population of the Milky Way (MW) dSphs, one of the most unique is the Sagittarius (Sgr) dSph. Kinematic analyses indicates that Sgr is a DM dominated galaxy. However, unlike almost all other dSphs, Sgr has an associated population of globular clusters, which are known gamma-ray emitters due to their millisecond pulsar populations~\cite{Fermi2010}. In particular the globular cluster at the center of Sgr, M54, is a well-motivated source for gamma-ray emission. Indeed, a gamma-ray source has been identified to be coincident with the spatial location of M54~\cite{20224FGL3DR}. It is unclear whether the observed gamma-ray emission is from the M54 millisecond pulsar  population or from DM annihilation within the Sgr halo~\cite{Evans:2022zno}. 

\par For the DM interpretation, it is crucial to accurately model the distribution of DM and the associated $\mathcal{J}$-factor, which is given by an integral over the DM phase space distribution. Therefore to accurately predict the flux due to annihilating DM, it is imperative to determine the spatial morphology of the DM halo and its mass density. Moreover, since Sgr is relatively nearby at a distance of $\sim26$ kpc from the Sun, it should have a relatively high $\mathcal{J}$-factor, and therefore flux, due to DM annihilation compared to other dSphs \citep{Pace:2019}. However, Sgr is located just below the Galactic Center as viewed from the Earth, at a location where the MW DM halo can contribute significantly to the observed DM annihilation signal. Furthermore, the DM halo of Sgr is likely spatially extended by several degrees on the sky, so that a standard point source or perhaps even a spherically shaped halo model for the gamma-ray emission may be an oversimplification.   

\par DM distributions in dSphs are best estimated using dynamical models under the assumption that the system is in dynamical equilibrium. In part because Sgr is in the process of tidal disruption, it has recently been suggested that there may be systematic issues in determining the $\mathcal{J}$-factor from equilibrium models~\cite{Venville:2023gye}. For this reason, it is optimal to turn to hydrodynamical simulations to understand the Sgr DM distribution. Recently, a sample of Sgr-like satellite galaxies in the Auriga simulations has been identified~\cite{Wang:2022kip}. These authors find that all of these Sgr analogues are DM dominated, and further that equilibrium modeling in their core region provides a faithful representation of the true DM distribution. Using the DM distribution of Sgr analogues from simulations more directly addresses the nature of the true morphology of the gamma-ray signal from annihilating DM. 

\par Sgr also represents a unique target for examining velocity-dependent DM annihilation. In this case, the $\mathcal{J}$-factor should also incorporate the full DM velocity distribution. Bounds on the annihilation cross section for p and d-wave models, in which the annihilation cross section, $\sigma_A v_{\rm rel}$, scales with DM relative velocity as $v_{\rm rel}^2$ and $v_{\rm rel}^4$, respectively, are much weaker than the limits from velocity-independent s-wave models. These different velocity scalings require different observational methods for extracting the signal~\cite{Smyth:2021bcp,Baxter:2022dpn}. The most stringent limits on the velocity-dependent cross section use the kinematic data from dSphs combined with simplified models for the DM velocity distribution~\cite{Boddy:2017vpe,Petac:2018gue,Boddy:2019qak}. These models for the velocity distribution may be compared to those from simulations~\cite{Board:2021bwj,Blanchette:2022hir,Piccirillo:2022qet}, which provide a more accurate representation of the DM relative velocity distribution. 

\par In this paper, we examine predictions for the DM annihilation signal from Sgr using analogues in the Auriga cosmological simulations. We provide predictions for the signals for both velocity-independent and velocity-dependent models. We consider, for the first time, the contribution to the DM annihilation signal from DM particles bound to the MW potential that spatially overlay with the Sgr analogues. For the purposes of this paper, we refer to this latter component as ``unbound'' particles, since they are not bound to the Sgr analogues. We demonstrate that the contribution from these unbound particles has a significant impact on the DM annihilation signals for the velocity-dependent models. We quantify the range of enhancement factors from this component and examine the implications for the sensitivity to velocity-dependent models using Fermi-LAT data. 

\section{Simulations and halo selections}
\label{sec:sims}

\subsection{The Auriga Simulations}

In this work we identify a set of Sgr analogues from the Auriga cosmological  simulations~\citep{Grand:2016mgo, Grand:2024}. The Auriga project~\citep{Grand:2016mgo} includes a suite of magneto-hydrodynamical zoom-in simulations of isolated MW mass halos. These halos were selected from a  $(100~{\rm cMpc})^3$ comoving volume periodic cube (L100N1504) from the parent DM-only simulations of the EAGLE project~\cite{Schaye:2014tpa, Crain:2015poa}. Cosmological parameters are adopted from the Planck-2015~\citep{Planck:2015fie} data: $\Omega_{m}=0.307$, $\Omega_b=0.048$, $H_0=67.77~{\rm km~s^{-1}~Mpc^{-1}}$. The simulations were performed using the moving-mesh code Arepo~\citep{Springel:2009aa}. They use a galaxy formation subgrid model which includes AGN and supernova feedback, star formation, black hole formation, metal cooling, and background UV/X-ray photoionisation radiation~\cite{Grand:2016mgo}. The simulations reproduce the observed stellar masses, sizes, star formation rates, rotation curves, and metallicities of MW mass galaxies at the present day.

We use the high resolution level 3 Auriga simulations in our study. At this resolution level, the DM particle mass is $m_{\rm DM} \sim 5\times 10^4~\Msun$, the baryonic particle mass is $m_b=5\times10^3~\Msun$, and the Plummer equivalent gravitational softening length is $\epsilon=184$~pc~\citep{Power:2002sw, Jenkins:2013raa}. The six MW analogues at the level 3 simulations studied in this work are Au6, Au16, Au21, Au23, Au24 and Au27.

\subsection{Identifying Sagittarius analogues}
\label{subsec:analogues}

We follow the study of ref.~\cite{Wang:2022kip} to identify the Sgr analogues in the Auriga simulations. There are multiple dwarf satellite galaxies associated with each of the six MW analogues. The dSphs are identified as Sgr analogues if their bound stellar mass is in the range of $10^{7.4}~\Msun < M_\ast <10^{8.5}~\Msun$ and their galactocentric distance is $\lesssim 60$~kpc, and they also have more than 6000 star particles. Dwarf galaxies whose geometrical minor axes and dynamical spin axes are strongly misaligned are excluded. For details on these selection criteria, see ref.~\cite{Wang:2022kip}. With these criteria, we find six Sgr analogues.

\begin{table*}
\makebox[\linewidth]{
	\centering
 \scalebox{0.84}{
	\begin{tabular}{|c|c|c|c|c|c|c|c|} 
            \hline
		Host &  Dwarf & $D$~[kpc] & $\log_{10}$($M_*$/M$_{\odot}$) & $\log_{10}$($M_{\rm tot}$/M$_{\odot}$) & $R_{\text{max}}$~[kpc] & Angular size [deg] & eccentricity\\
				\hline
    Au16 & 9 & 28.08 & 7.933 & 8.810 & 1.957 & 3.945 & 0.65 \\
    Au21 & 10 & 44.70 & 8.446 & 8.780 & 1.078 & 1.395 & 0.90 \\
    Au23 & 4 & 49.01 & 8.297 & 8.851 & 1.583 &  1.884 & 0.08 \\
    Au23 & 7 & 38.11 & 8.047 & 8.790 & 1.654 & 2.502 & 0.67 \\
    Au24 & 24 & 60.13 & 7.503 & 8.629 & 2.970 & 2.788 & 0.62 \\
    Au27 & 25 & 35.91 & 7.598 & 8.170 & 0.852 & 1.390 & 0.66 \\
    \hline
    Sgr & & $26.4$ & $8.6$ & & $\gtrsim 1$ & $\gtrsim 1 2.2$ & 0.65 \\
    \hline 
	\end{tabular}
 }}
	\caption{Summary of the properties of the six simulated Sgr-like systems in Auriga. From left to right, the columns list the Auriga halo ID of the MW analogue hosts, the dwarf ID of the Sgr analogues, the distance of the observer to the dwarf, the present-day bound stellar mass, total present-day bound mass, radius $R_{\rm max}$, the angular size, and the eccentricity of the Sgr analogues. The angular size is calculated using the distance from the halo center to the Sun. The bottom row gives the properties of the observed Sgr dwarf. 
 \label{tab:wang_tab}}
\end{table*}

Table~\ref{tab:wang_tab} lists the properties of the six selected Sgr analogues. The columns of the table from left to right list the Auriga halo ID of the MW analogue hosts, the dwarf ID of the Sgr analogues, distance of the Sgr analogues to the observer, the bound stellar mass of the Sgr analogue, the total (DM and stars) bound mass of the Sgr analogue, the radius, $R_{\rm max}$, at which the circular velocity of the Sgr analogue reaches its maximum value,  the angular size of the Sgr analogues as seen from the Solar position, and the eccentricity of the Sgr analogues. In the bottom row we present the properties of the observed Sgr dwarf.

\begin{figure}[t!]
\centering
\includegraphics[width=1\textwidth]{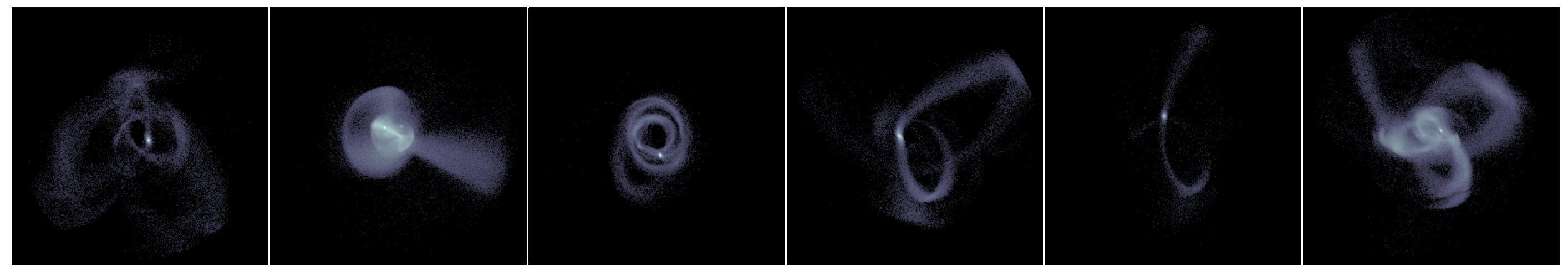}
\caption{
    Stellar mass maps of the tidal debris associated with each Sgr analogue identified in Auriga.
    From left to right are Sgr(9, 10, 4, 7, 24, 25) as denoted in table~\ref{tab:wang_tab}, with each 500 kpc per side panel centred on the MW host.
    As extreme examples, Sgr 10 is incredibly massive ($M_\ast \sim 10^{9.5}$~M$_\odot$) and on a very radial orbit $(e\sim0.9)$, while Sgr 4 is on a nearly circular orbit (with eccentricity $e\sim0.1$).
    In terms of stellar mass and orbital trajectory, Sgr~9 and Sgr~7 are the closest matches to the real Sgr (see Section \ref{subsec:analogues} for further discussion).
} \label{fig:debris-maps}
\end{figure}

While the definition of Sgr analogues in ref.~\cite{Wang:2022kip} relies on the stellar mass bound to the progenitor at present day, it can be useful to also consider other properties like morphology of tidal debris and orbital trajectory.
Figure \ref{fig:debris-maps} shows the distribution of stellar mass for each Sgr analogue, including mass that has been lost after accreting onto the MW host.
It is clear that the selection criteria results in a variety of possible morphologies, largely driven by the total mass at infall and the orbit of the analogues.
Sgr has an estimated total stellar mass (prior to infall) of $M_\ast \sim 10^{8.3}~\Msun$ \citep{Niederste-Ostholt:2010} and eccentricity $e \equiv (r_
\text{apo} - r_\text{peri})/ (r_
\text{apo} + r_\text{peri}) \sim 0.65$ \citep{Vasiliev:2021}.
Of the analogues listed in table~\ref{tab:wang_tab}, Sgr 9 and Sgr 7 fall into a reasonable range of matches, with stellar masses prior to infall of $10^{8.5}~\Msun < M_\ast < 10^{8.7}~\Msun$ and eccentricities $0.6 < e < 0.75$. Note that these infall stellar masses are different from the present-day stellar masses quoted in table~\ref{tab:wang_tab}. An analysis of the full population of stellar streams in the Auriga simulations is forthcoming~\citep{Riley:2024,Shipp:2024}.

\subsection{Finding the Solar positions}

In the simulations, the Solar position is not determined a priori. For each MW-Sgr system, we choose the Solar position such that the position of the Sgr analogue matches its correct on-sky location ($l =5.569^\circ$, $b =-14.166^\circ$), with the galactic center  at $l =0^\circ$, $b =0^\circ$. To determine the Solar position, we use a temporary spherical coordinate system centered on the galactic center and sample over spherical shells of  0.01 kpc radial width that are at a galactocentric distance ranging from $7 - 9$~kpc. Within each spherical shell centered on the galactic center, the position is checked for 500 points by varying the azimuthal angle between 0 and $2\pi$, and 250 points by varying the inclination angle between 0 to $\pi$. Since there is no well-defined disk for the host halo galaxy, we do not enforce a constraint to keep the Sun in the galactic disk. We only constrain its distance to the galactic center. For each system, we find the Sgr analogue positions are within 0.2 degree of the targeted latitude and longitude. The resulting distances between the best Solar position and its associated Sgr analogue are given in the third column of table \ref{tab:wang_tab}. We find that the range of circular velocities, $v_c = \sqrt{GM(<r)/r}$ at the Solar position for each host analogue is in the range $[233 - 270]$~km/s, which is consistent with the observed value~\cite{2017MNRAS.465...76M}. The  mass, $M(<r)$, is computed as the sum of the masses of dark matter, gas, and stars within radius $r$ from the center of the host halos.

\section{Dark matter density and velocity distributions}

In this section we discuss the DM density profiles and pair-wise relative velocity distributions of the Sgr analogues. The DM density profile is an important quantity which enters in the calculations of the predicted DM annihilation signal and the $\mathcal{J}$-factor. The DM relative velocity distribution is relevant for the case of velocity-dependent DM annihilation models and enters the $\mathcal{J}$-factor through the velocity dependence of the annihilation cross section~\cite{Robertson:2009bh,Ferrer:2013cla, Zhao:2017dln,Boddy:2018ike,Lacroix:2018qqh,Boddy:2019wfg,McKeown:2021sob}. 

For each Sgr analogue, we extract the position vector, ${\bf x}$, and the velocity vector, ${\bf v}$, of the bound and unbound DM particles, with respect to the center of that subhalo. The bound DM particles are those identified by the SUBFIND algorithm~\cite{Springel:2000qu}. The unbound DM particles are those belonging to the host MW analogue, but spatially coinciding with the Sgr analogue, which we define as lying within a spherical region of radius $R_{\rm max}$ of the analogue. The phase-space distribution, $f(\textbf{x}, \textbf{v})$, is defined such that $f(\textbf{x}, \textbf{v})~d^3\textbf{x}~d^3\textbf{v}$ is the mass of DM particles within a phase space volume $\textbf{x}+d^3\textbf{x}$ and $\textbf{v}+d^3\textbf{v}$. The probability distribution of DM velocities at a position ${\bf x}$ can be written as
\begin{equation}
P_\textbf{x} (\textbf{v}) = \frac{f(\textbf{x},\textbf{v})}{\rho(\textbf{x})},
\label{eq:DM-vel-prob}
\end{equation}
where the DM density at $\textbf{x}$ is given by $\rho(\textbf{x}) = \int f(\textbf{x},\textbf{v})d^3\textbf{v}$.

We can then write an expression for the probability distribution of relative velocities at a position $\textbf{x}$ for a pair of DM particles with velocities ${\bf v}_1$ and ${\bf v}_2$,
\begin{equation}
P_\textbf{x}(\textbf{v}_{\rm rel}) = \int P_\textbf{x}(\textbf{v}_1={\bf v}_{\rm cm}+{\bf v}_{\rm rel}/2) P_\textbf{x}(\textbf{v}_2 = {\bf v}_{\rm cm}-{\bf v}_{\rm rel}/2)~d^3\textbf{v}_{\rm cm}.
\label{eq:P-x-vrel}
\end{equation}
where $\textbf{v}_{\rm cm}$ is the center-of-mass velocity of the two particles and $\textbf{v}_{\rm rel} \equiv \textbf{v}_2 - \textbf{v}_1$ is their relative velocity.

From the DM relative velocity distribution, $P_\textbf{x}(\textbf{v}_{\rm rel})$, we can find the DM relative velocity modulus (i.e. speed) distribution
\begin{equation}
P_\textbf{x}(|\textbf{v}_{\rm rel})| =v_{\rm rel}^2 \int P_\textbf{x}(\textbf{v}_{\rm rel})\,d\Omega_{\bf v_{\rm rel}}\, ,
\end{equation}
where $d\Omega_{\bf v_{\rm rel}}$ is an infinitesimal solid angle along the direction $\textbf{v}_{\rm rel}$. 

\begin{figure}[t]
\centering
\includegraphics[width=1\textwidth]{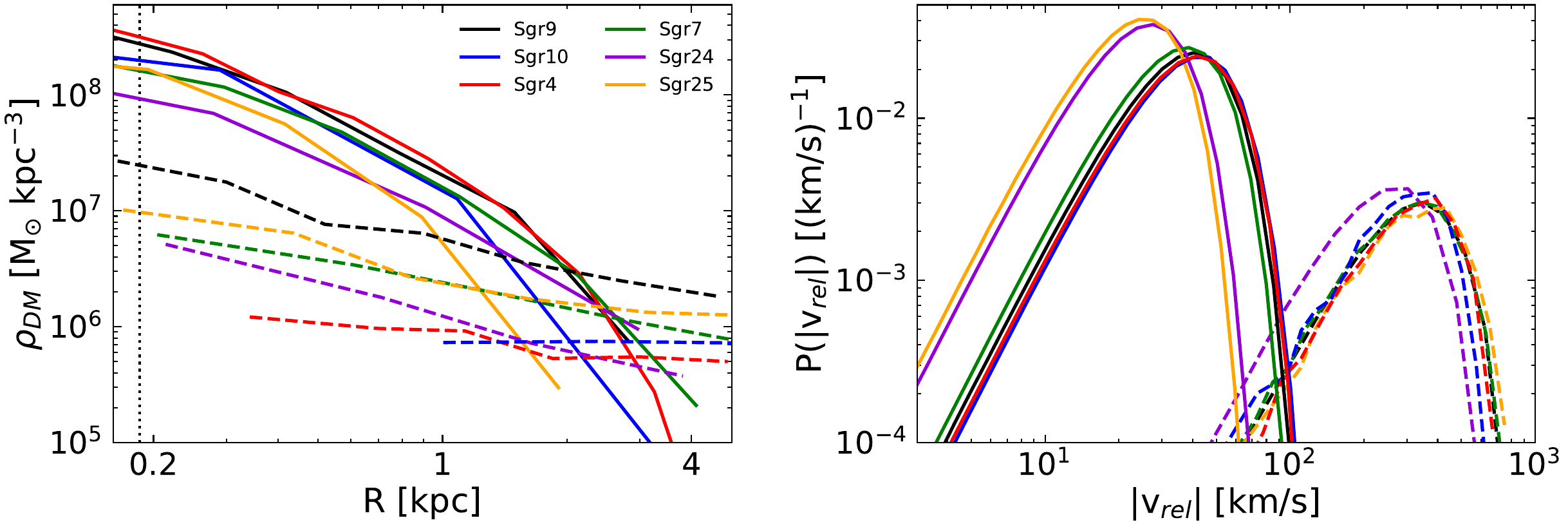}
\caption{
DM density profiles (left panel) and DM relative speed distributions (right panel) for the six selected Sgr analogues. Solid curves correspond to the DM particles bound to the Sgr analogues, and dashed curves correspond to DM particles belonging to the smooth halo of the MW analogues that are spatially coincident with the Sgr system and are thus considered unbound. For the dashed curves in the right panel, only unbound particles within radius $R_{\rm max}$ of each Sgr analogue are considered. The vertical dotted line in the left panel indicates the gravitational softening length of $\epsilon=185$~pc. } \label{fig:DMdist}
\end{figure}

In the left panel of figure~\ref{fig:DMdist} we present the DM density profiles for our selected Sgr analogues. We extract the spherically-averaged DM density profiles for each Sgr analogue from the DM mass enclosed within consecutive spherical shells of variable radial bin widths from the center of the subhalo. Particles are binned such that the minimum number of particles per bin is 20. The DM density profiles are presented for two different DM populations. The solid curves correspond to the density profiles of DM particles bound to each subhalo, as identified by the SUBFIND algorithm~\cite{Springel:2000qu}. The dashed curves, on the other hand, show the density profiles for DM particles belonging to the smooth halo of the MW analogues, which are spatially coincident with the Sgr analogue and are considered unbound to the Sgr analogues. The vertical dotted line indicates the gravitational softening length of $\epsilon=185$~pc. At the center of the Sgr analogues, the density of the unbound DM particles is roughly an order of magnitude smaller than the bound DM particles. These two densities become comparable in the outer regions. Notice that in the central regions of Sgr 4 and Sgr~10, there are too few unbound DM particles, so the density curves start at a larger radius, such that the inner most radial bin contains at least 20 particles.

The right panel of figure~\ref{fig:DMdist} shows the relative DM speed distributions for our selected Sgr analogues for the bound (solid curves) and unbound (dashed curves) DM particles. Unbound DM particles used for this plot are those belonging to the smooth halo of the MW analogues, which are within the $R_{\rm max}$ of each Sgr analogue. The DM relative speed distributions are extracted considering non-uniform velocity bins such that there are at least 20 particles per bin.  
As it can be seen from the figure, the DM relative speed distributions of the unbound DM particles peak at much higher speeds compared to the bound DM particles. This is expected since the average relative speeds of the DM particles of the host MW-like halos at the position of the Sgr analogues (i.e.~$[20-60]$~kpc from the host halo's center) are larger than the average relative speeds of the DM particles bound to the Sgr analogues.

The unbound DM particles which reside in the spatial region of specific subhalos are generally not included in the analysis of DM annihilation signals from those subhalos or dSph galaxies, perhaps due to their low densities compared to bound particles. However, due to the large difference between their peak relative speeds with the peak relative speeds of the bound DM particles, they are expected to have a significant impact on the velocity-dependent $\mathcal{J}$-factors, as we will discuss in the sections below. We therefore include these unbound particles in the analysis of the $\mathcal{J}$-factors in this work for the first time.

\section{J-factors} 
\label{sec:jfactor}

\par In this section we discuss the calculation of the $\mathcal{J}$-factors for both velocity-independent and velocity-dependent DM annihilation models. We emphasize the differences when including the unbound DM particles in the calculation, as described above. We begin by presenting the results of the full calculation, then compare to the case in which the unbound DM particles are excluded. 

\subsection{Including both bound and unbound particles}

The $\mathcal{J}$-factors for velocity-dependent annihilation models depend  on both the DM density profile and the DM pair-wise, or relative velocity distribution. We follow closely the formalism in refs.~\cite{Board:2021bwj, Blanchette:2022hir} to compute the $\mathcal{J}$-factors for each of the annihilation cross section models. In this section, we include both the  DM particles bound to the Sgr analogues and the unbound DM particles which reside in the spatial region of the Sgr analogues, in the calculation of the $\mathcal{J}$-factors.

The DM annihilation cross section, $\sigma_A$, averaged over the relative velocity distribution at position, $\textbf{x}$, in a halo  is given by
\begin{equation}
\langle \sigma_A v_{\rm rel} \rangle (\textbf{x}) = \int d^3 \textbf{v}_{\rm rel} P_\textbf{x}(\textbf{v}_{\rm rel})(\sigma_A v_{\rm rel}).
\label{eq:annihilation-cross-section}
\end{equation}
In general, $\sigma_A v_{\rm rel}$ can be parametrized as $\sigma_A v_{\rm rel} = (\sigma_A v_{\rm rel})_0 (v_{\rm rel}/c)^n$, and depends on the relative DM velocity. $(\sigma_A v_{\rm rel})_0$ is the velocity-independent component of the annihilation cross section, and $n$ depends on the specific DM annihilation model. We consider the following models: $n=0$ (s-wave annihilation), $n=2$ (p-wave annihilation), $n=4$ (d-wave annihilation), and $n=-1$ (Sommerfeld-enhanced annihilation).

The effective $\mathcal{J}$-factor is defined as~\cite{Boddy:2019wfg, Board:2021bwj},
\begin{equation} 
{\mathcal J} (\theta) = \int d \ell \, \frac{\langle \sigma_A v_{\rm rel} \rangle}{(\sigma_A v_{\rm rel})_0}  \left[\rho (r(\ell, \theta))\right]^2 = \int d \ell \int d^3 {\bf v}_{\rm rel} P_{{\bf x}} ({\bf v}_{\rm rel}) ~\left(\frac{{v}_{\rm rel}}{c}\right)^n~ \left[\rho (r(\ell, \theta))\right]^2\, ,
\label{eq:Jfactor}
\end{equation} 
where $\ell$ is the line of sight distance from the Sun to a point in the Sgr analogue, $\theta$ is the opening angle between the line of sight $\ell$ and the distance $D$ from the Sun to the center of the Sgr analogue, and $r^2(\ell, \theta) = \ell^2 + D^2 - 2\ell D \cos{\theta}$ is the square of the radial distance measured from the center of the Sgr analogue. This equation assumes that the Sgr analogue is spherically symmetric.

The expected gamma-ray flux from DM annihilation is, in general, proportional to ${\mathcal J}$ and can be written as,
\begin{equation}
\frac{d \Phi_\gamma}{dE} = \frac{\left(\sigma_A v_{\rm rel}\right)_0}{8 \pi m_\chi^2}\frac{dN_\gamma}{dE} \,\mathcal{J},
\label{eq:flux}
\end{equation}
where $m_\chi$ is the DM particle mass, and $dN_\gamma/dE$ is the gamma-ray energy spectrum produced per annihilation.

To compute the effective $\mathcal{J}$-factor for a Sgr analogue, we need to include three separate components. These components are: (1) the DM particles within the Sgr analogue (bound + unbound), (2) the DM particles belonging to the smooth halo along the line-of-sight to the Sgr analogue, and (3) the DM particles bound to subhalos along the line-of-sight to the Sgr analogue.

The $\mathcal{J}$-factors are computed separately for the smooth halo and subhalos. For the smooth halo component, the local DM density around each DM particle is estimated using a Voronoi tessellation. Then the DM relative velocity distribution is computed at each point on a spherical mesh using the nearest 500 DM particles~\cite{Piccirillo:2022qet}. These values are then interpolated to obtain an estimate for the DM relative velocity distribution at every point in the smooth halo. The $\mathcal{J}$-factors are then computed by integrating the contribution from all DM particles along the line-of-sight to the Sgr analogue. For subhalos (including the Sgr analogues), the local DM density is found using the best fit Einasto density profile within $R_{\rm max}$ and a Voronoi tessellation outside of $R_{\rm max}$. The density profile is computed such that each shell has 100 DM particles. The DM relative velocity distribution is also calculated in non-uniform consecutive spherical shells from the center of the subhalo, where each shell contains 100 DM particles. This allows us to capture the radial dependence of the DM relative velocities in the subhalo. The resulting velocity distribution is then used to compute the effective $\mathcal{J}$-factors using eq~\eqref{eq:Jfactor}. \

\begin{figure}[t]
\centering
\includegraphics[width=1\textwidth]{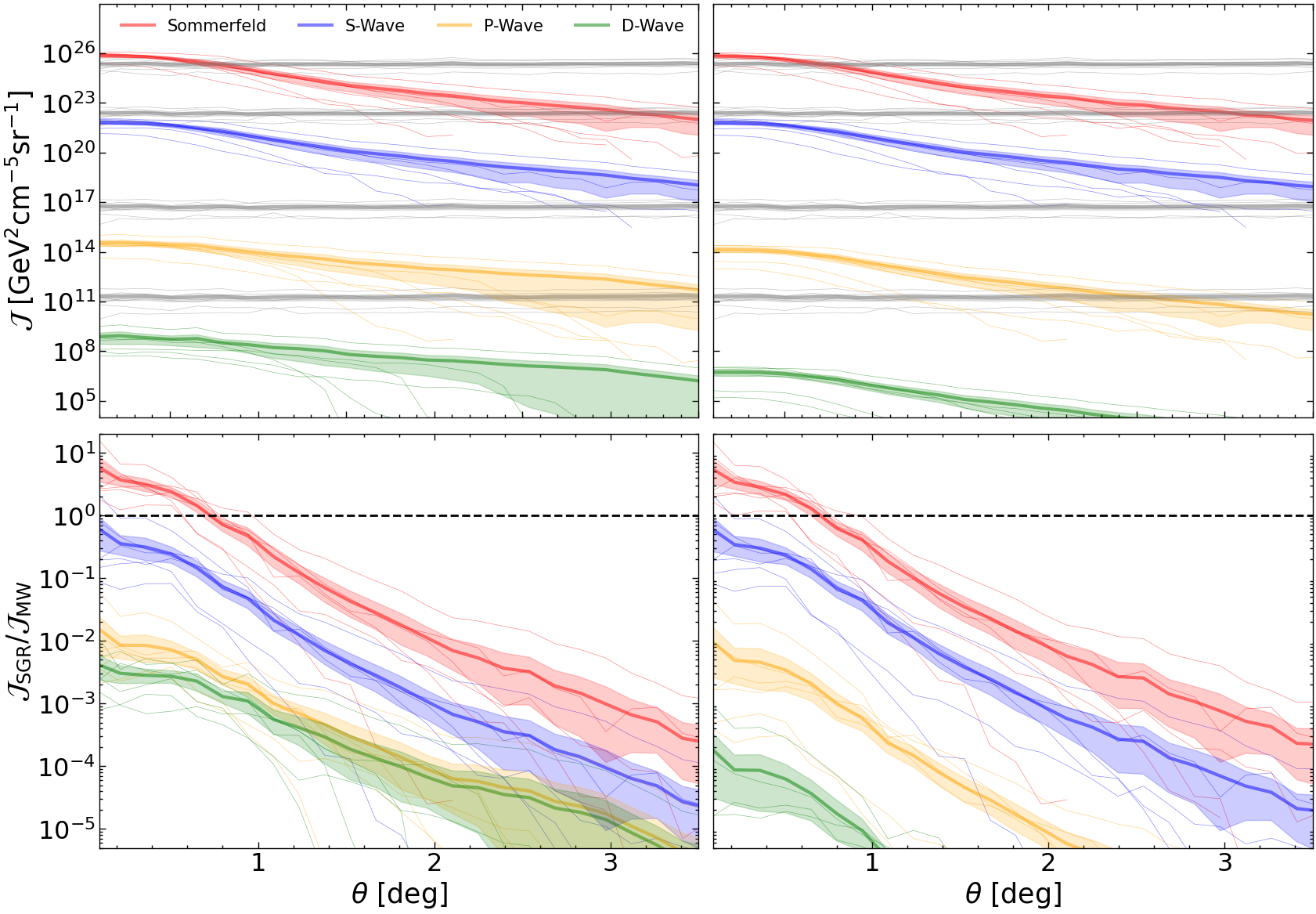}
\caption{Top: $\mathcal{J}$-factor profiles for the Sgr analogues (coloured lines) and the smooth component of their host halo along the line-of-sight (grey lines). Bottom: Ratio of the $\mathcal{J}$-factors for each Sgr analogue and the smooth component of its host MW-like halo along the line-of-sight to the Sgr analogue (coloured lines). The  Sgr $\mathcal{J}$-factors in the left panels are calculated with bound and unbound DM particles included as part of the Sgr analogues, while in the right panels, only particles bound to the Sgr analogues are included. In all panels, the thin coloured lines show the results for each Sgr analogue, while the thick coloured lines correspond to the average  across all Sgr analogues,  with the shaded band indicating the standard error around the mean. The thin and thick grey lines in the bottom panel correspond to the $\mathcal{J}$-factors for the smooth MW-like host halos and the average $\mathcal{J}$-factor across all host halos, respectively. The red, blue, orange, and green coloured lines show the results for the Sommerfeld, s-wave, p-wave, and d-wave DM annihilation models, respectively. 
}
 \label{fig:J}
\end{figure}

The top left panel of figure~\ref{fig:J} shows the radial $\mathcal{J}$-factor profiles of the Sgr analogues including both bound and unbound DM particles (coloured lines) and the smooth component of their host MW-like halo  along the line-of-sight to the Sgr analogue (grey lines). $\mathcal{J}$-factors are computed in evenly spaced angular bins on the sky centered on the Sgr analogues. Note that the subhalo contribution along the line-of-sight to the Sgr analogue is negligible, and we have therefore not included it in the calculation of the $\mathcal{J}$-factor of the host MW-like halo in this figure. The bottom left panel of the figure shows the ratio of the radial $\mathcal{J}$-factor profiles for each Sgr analogue and the smooth halo contribution from its host halo along the line-of-sight. In both the top and bottom panels, the thin coloured lines show the results for each Sgr analogue, while the thick coloured lines correspond to the average  across all Sgr analogues, with the shaded band indicating the standard error around the mean. The thin and thick grey lines in the top panel correspond to the $\mathcal{J}$-factors for the smooth MW-like host halos and the average $\mathcal{J}$-factor across all host halos, respectively. The red, blue, orange, and green coloured lines show the results for the Sommerfeld, s-wave, p-wave, and d-wave DM annihilation models, respectively. 

\begin{figure}[t]
\centering
\includegraphics[width=1\textwidth,height=0.5\textwidth]{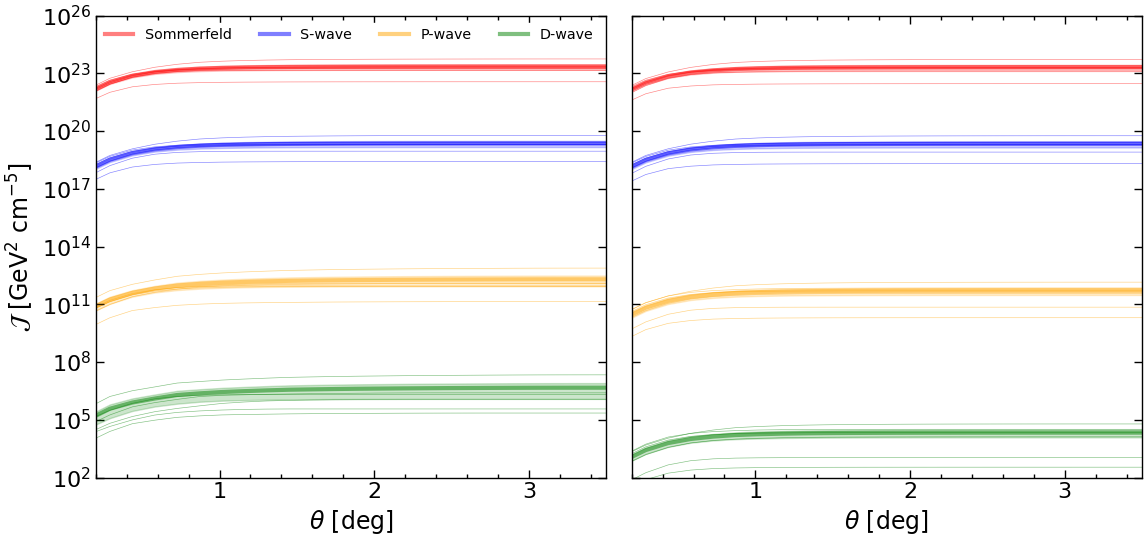}
\caption{Integrated $\mathcal{J}$-factor profiles (eq.~\eqref{eq:integrated-J}) for the Sgr analogues. In the left panel the $\mathcal{J}$-factor are calculated with the bound and unbound DM particles, while in the right panel
only the bound DM particles were included in the calculation. In both panels, the thin coloured lines show the results for each Sgr analogue, while the thick coloured lines correspond to the average across all Sgr analogues. The shaded bands indicate the standard error around the mean. The red, blue, orange, and green coloured lines show the results for the Sommerfeld, s-wave, p-wave, and d-wave DM annihilation models, respectively.
} 
\label{fig:integrated} 
\end{figure}

An important quantity for DM indirect detection is the $\mathcal{J}$-factor integrated over solid angle, which is given by
\begin{equation}
 \widetilde{\mathcal{J}}(\theta)=2\pi \int_0^\theta \mathcal{J}(\theta') \sin \theta' d\theta' . 
 \label{eq:integrated-J}
\end{equation}
In the left panel of figure~\ref{fig:integrated}, we show the integrated $\mathcal{J}$-factor profiles for the Sgr analogues including both bound and unbound DM particles. Same as in figure~\ref{fig:J}, the thin lines show the results for different Sgr analogue, while the thick lines indicate the average  across all Sgr analogues. The shaded bands show the standard deviation around the mean. The red, blue, orange, and green lines correspond to the Sommerfeld, s-wave, p-wave, and d-wave models, respectively. The integrated $\mathcal{J}$-factors for the s-wave reach values at large $\theta$ which are consistent with the $\mathcal{J}$-factor computed for the observed Sgr~\cite{Evans:2022zno}.

\par
\begin{figure}[t]
\centering
\includegraphics[width=1\textwidth,height=0.45\textwidth]{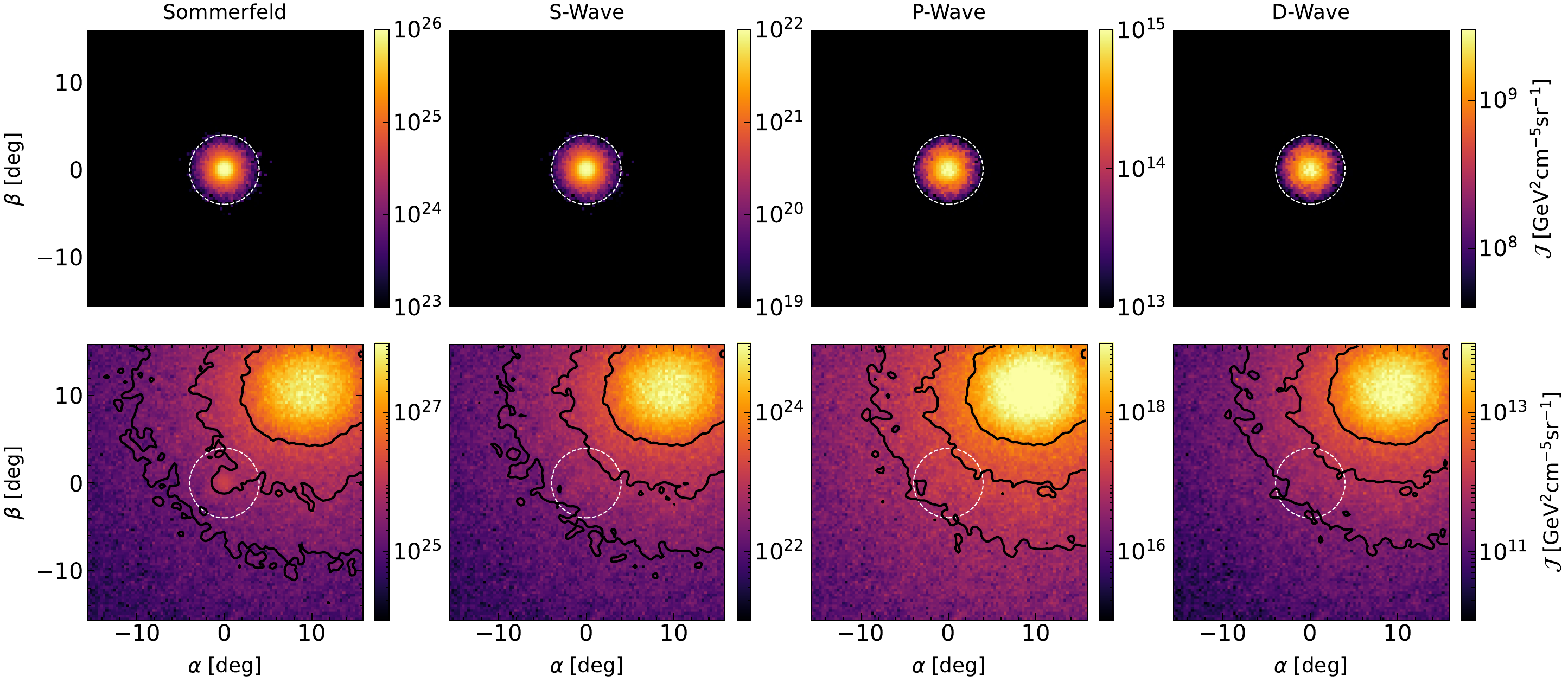}
\caption{$\mathcal{J}$-factor maps for a region of interest centered on the Sgr 9 analogue for the Sommerfeld, s-wave, p-wave, and d-wave models. The top panels show only the contribution from the Sgr 9 analogue, including both bound and unbound DM particles. The bottom panels show the contribution from the entire region, including the MW foreground emission. $\alpha$ and $\beta$ are the polar and azimuthal angles in a randomly defined spherical coordinate system centered on the Sgr analogue, with the $z$-axis pointing from the Sun's position to the Sgr analogue. The dashed white circle specifies the angular extension of the $R_{\rm max}$ of the Sgr source. The bright component in the upper right corner of each map in the bottom panels is the emission from DM annihilation from the Galactic center. Only in the Sommerfeld case is the emission from the Sgr source clearly visible above the diffuse emission from the MW. In the bottom panels, the black contours are placed at fractions of the maximum pixel value given by 0.02, 0.005 and 0.0015.
\label{fig:jfactormaps} }
\end{figure}

 Figure~\ref{fig:jfactormaps} shows the $\mathcal{J}$-factor maps for a region around the Sgr 9 analogue, for the different velocity-dependent DM annihilation models. For each panel we use a pixel size of 0.3 degree on each side. The top panels show the $\mathcal{J}$-factor maps for only the Sgr 9 analogue including bound and unbound particles, while the bottom panels show the $\mathcal{J}$-factor maps for Sgr 9 once the contribution from the MW smooth halo and other subhalos in the region are added in. We choose to highlight Sgr 9 as it is the Sgr analogue with the largest angular size, and is therefore most easily observed in the maps. As shown in the figure, the MW diffuse signal from the smooth halo dominates the source signal in all cases but the Sommerfeld case, where the velocity dependence goes as $1/v$. For the Sommerfeld case, we see that the source is clearly visible above the MW emission. For the s, p and d-wave cases, the source is not visible above the MW emission.

\subsection{Including only bound particles}
\par It is interesting to compare the true $\mathcal{J}$-factors computed from the bound and unbound DM particles in the Sgr analogues 
to the hypothetical case in which we only consider the emission from DM particles that are classified as bound to the Sgr analogues. For this comparison, we exclude the particles from the $\mathcal{J}$-factor calculation that spatially overlap with the Sgr analogues, but are bound to the MW host halos and not to the Sgr analogues. We then compare the results to the $\mathcal{J}$-factors calculated using the full simulation. The spherically-averaged  $\mathcal{J}$-factors as a function of angle from the center of the Sgr analogues are shown in the right panels of figure~\ref{fig:J}, where only the bound particles to the Sgr analogues are included. Comparing to the left panel, the largest relative enhancements due to the unbound DM particles in the Sgr region come from the d-wave component, although the total emission is still well below that of the MW emission in this case. It is interesting to note that for the case of  p and d-waves, there is a larger scatter when including the unbound DM particles. For the Sommerfeld case, the effect of the high speed unbound DM particles is suppressed due to the $1/v$ dependence of this model, and thus they have minimal effect on the $\mathcal{J}$-factors. The effect is also minimal for s-wave since there is no velocity dependence and the density of unbound DM particles is low. Similar results are found for the integrated $\mathcal{J}$-factors when comparing the left panel of figure~\ref{fig:integrated} to the right panel, which only considers the bound DM particles in the calculation.

\begin{table*}[t]
	\centering
	\begin{tabular}{|c|c|c|c|c|c|} 
            \hline
		Host name &  Dwarf & Sommerfeld & S-wave & P-wave & D-wave\\
				\hline
    Au16 & 9 & 1.28 & 1.22 & 16.79 & 1379.01  \\
    Au21 & 10 & 1.06 & 1.03 & 1.68 & 15.76 \\
    Au23 & 4 & 1.09 & 1.05 & 2.85 & 103.19 \\
    Au23 & 7 & 1.15 & 1.07 & 5.15 & 342.22 \\
    Au24 & 24 & 1.49 & 1.50 & 29.83 & 3878.96 \\
    Au27 & 25 & 1.24 & 1.11 & 15.12 & 2412.44 \\ 
    \hline
	\end{tabular}
	\caption{Modifications to the $\mathcal{J}$-factors of the Sgr analogues due to  the  unbound DM particles. Each entry shows the ratio of the $\mathcal{J}$-factor for the full simulation results (including both the bound and unbound DM particles) to the case in which only the DM particles bounds to the Sgr analogues are considered. The ratios specify the average enhancements over all pixels incorporating each Sgr analogue. Note that there is a considerable enhancement for p and d-waves, while there is 
 essentially no change in the Sommerfeld and s-wave cases. }
 \label{tab:modifications}
\end{table*}

\begin{figure}[t]
\centering
\includegraphics[width=1\textwidth,height=0.5\textwidth]{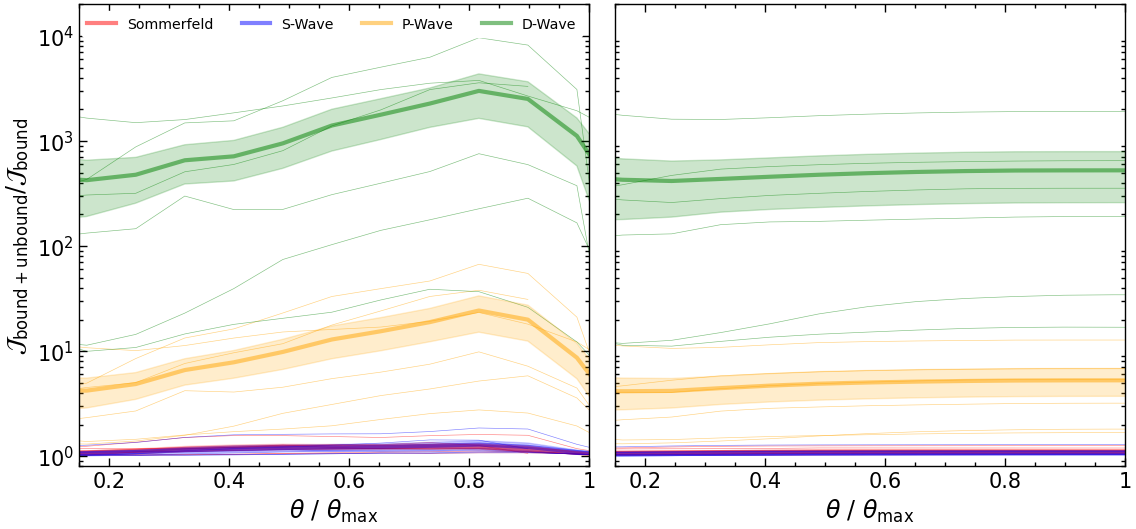}
\caption{Ratio of the $\mathcal{J}$-factors of the Sgr analogues calculated with both bound and unbound DM particles to those calculated with only the bound DM particles. In the left panel, the ratios are shown for the differential $\mathcal{J}$-factors, while in the right panel they are shown  for the integrated $\mathcal{J}$-factors. In both panels, the thin coloured lines show the results for each Sgr analogue, while the thick coloured lines correspond to the average across all Sgr analogues, with the shaded band indicating the standard error around the mean. The red, blue, orange, and green coloured lines show the results for the Sommerfeld, s-wave, p-wave, and d-wave models, respectively.}
\label{fig:integrated_differential} 
\end{figure}

\par The enhancement factors when including the unbound DM particles in the calculation of the $\mathcal{J}$-factors of the Sgr analogues for the four DM annihilation models are shown in table~\ref{tab:modifications}. The values shown are the average enhancement factors over all pixels that make up each respective Sgr analogue. Including the unbound DM particles can increase the $\mathcal{J}$-factors of the Sgr analogues by over an order of magnitude for the p-wave model, and by over three orders of magnitude for the d-wave model. For the Sommerfeld and s-wave models, 
the $\mathcal{J}$-factors minimally change.

\par The angular dependence of the enhancement factors can be seen in figure~\ref{fig:integrated_differential}, which shows the ratio of the $\mathcal{J}$-factors calculated with the bound and unbound DM particles to those calculated with only the bound DM particles. We define $\theta_{\rm max}$ as the angular extent of the subhalo at radius $R_{\rm max}$. The differential $\mathcal{J}$-factor profile ratios are shown in the left panel and the integrated $\mathcal{J}$-factor profile ratios are shown in the right panel. The line colors, styles, and shaded bands are the same as in figures~\ref{fig:J} and \ref{fig:integrated}. While the enhancement factors are significant for p and d-wave models for all angles and for both the differential and integrated $\mathcal{J}$-factors, they remain negligible for the Sommerfeld and s-wave models. The ratios of the differential $\mathcal{J}$-factors in the left panel peak slightly below $R_{\rm max}$, due to the decreasing contributions from the lines-of-sight towards the edge of the subhalo.

\section{Gamma-ray data and model fitting}
\label{sec:gammarays}

\par Recent analyses has found a gamma-ray source that is coincident with the spatial location of M54, located at the center of Sgr~\cite{20224FGL3DR}. This source has a spectral energy distribution that is consistent with either a millisecond pulsar interpretation or a DM interpretation~\cite{Evans:2022zno}. For the DM interpretation, the best-fitting DM mass is approximately 30 GeV assuming full annihilation into $b \bar b$. While the source is consistent with having a small extension of $\lesssim 1^\circ$, the
point source interpretation is mildly favored over an extended source template. Given the faint nature of the source and the proximity to diffuse Galactic foreground emission, larger extensions of $\gtrsim 1^\circ$ are difficult to model and establish evidence of emission from. Interestingly, noting the results from figure~\ref{fig:J}, the DM emission from the Sgr analogues extend to greater than $1^\circ$. Therefore, in order to test our extended templates generated above against the data, we must carefully account for the extension and possible confusion from foreground emission.

\par Using the $\mathcal{J}$-factor maps generated from the simulations as templates, we perform an extended source binned likelihood analysis on the Sgr region. 
To analyze the data we use the Fermitools 2.0.8.\footnote{https://github.com/fermi-lat/Fermitools-conda/wiki}. 
We utilize~\texttt{FermiPy}~\citep{Wood:2017yyb}, which is a python-based software package that automates the tools for Fermi-LAT source analysis. 
For our data selection, we use~\texttt{Pass 8 SOURCE}-class photon events with the corresponding instrument response functions,~\texttt{P8R3}\textunderscore\texttt{SOURCE}\textunderscore\texttt{V3}. 
We select both \texttt{FRONT} and \texttt{BACK} converting events (evclass=128 and evtype=3), with energies in the range of 300~MeV to 500~GeV. 
For our primary analysis, we exclude photons with energies below 300~MeV in order to avoid complications associated with the broader point spread function~\citep{20224FGL3DR}.
We use approximately 15.5 years of data, corresponding to mission elapsed times between 239557417~s and 728225229~s. 
We apply the recommended \texttt{(DATA\char`_QUAL>0)\&\&(LAT\char`_CONFIG==1)} filter to ensure quality data and a zenith cut of $z_{\rm max} = 90^\circ$ to filter background gamma-ray contamination from the Earth's limb. 

\par We consider a $15^\circ \times 15^\circ$ ROI centered on M54/Sgr. 
For our likelihood maximization, we take a 0.1$^\circ$ angular pixelation and use the \texttt{MINUIT} optimizer method within \texttt{gtlike}. 
We use an input source model that includes all sources in the 4FGL-DR3 catalog~\citep{20224FGL3DR}~out to a square of $20^\circ \times 20^\circ$. 
Including sources beyond the ROI ensures that sources on the edge of the ROI are properly modeled. For the interstellar emission model we use the recommended \texttt{gll}\_\texttt{iem}\_\texttt{v07.fits}, and for the isotropic emission we use \texttt{iso}\_\texttt{P8R3}\_\texttt{SOURCE}\_\texttt{V3}\_\texttt{V1.txt}. 
To report the significance of our sources, we report the TS value obtained upon running \texttt{gta.sed}, which takes the source following the procedure outlined above and then performs a bin-by-bin likelihood fit assuming a power law spectralization per bin with an index set to $\Gamma = -2$.
Finally, for all TS values reported we assume a DM annihilation spectrum, as defined by~\texttt{DMFitFunction} provided in~\text{FermiPy} assuming annihilation via the $b\bar{b}$ channel. 
For this spectrum, we fit the $\mathcal{J}$-factor to the total integrated $\mathcal{J}$-factor for a given source map, fix the mass to the best-fitting mass found in ref.~\cite{Evans:2022zno} of 29.6 GeV, and allow $\langle \sigma v \rangle$ to float.

\par  The flux of annihilating DM from the MW halo is larger than that of Sgr for s-wave, and dominates the Sgr source for p and d-waves. We obtain $\mathcal{J}_{\text{Sgr}} / \mathcal{J}_{\text{MW}} \sim 0.1 - 1$ for s-wave, $\mathcal{J}_{\text{Sgr}} / \mathcal{J}_{\text{MW}} \sim 0.003 - 0.03$ for p-wave, and $\mathcal{J}_{\text{Sgr}} / \mathcal{J}_{\text{MW}} \sim 0.001 - 0.005$ for d-wave, where $\mathcal{J}_{\text{MW}}$ denotes the MW $\mathcal{J}$-factor integrated over the area of the target. For the Sommerfeld case, the Sgr source flux is always larger than the MW halo emission, with $\mathcal{J}_{\text{Sgr}} /\mathcal{J}_{\text{MW}} \sim 2 - 7$. When fitting both the Sgr source and the MW diffuse over the entire ROI, we find an overall poor fit quality, because we are modeling the diffuse emission over such a large region of sky. This likely means that an appropriate handling of the MW DM emission requires unique templates for the diffuse MW emission arising from cosmic ray interactions with gas and the interstellar radiation field, which is beyond the scope of this work.
Therefore, for the analysis in this section we model only the emission associated with the Sgr source.
In this way, we can compare qualities of the source independent of the MW diffuse emission, such as the effect of including the unbound DM particles in our templates.

\par To model M54/Sgr in gamma-rays based on our simulations results, using the results from figure~\ref{fig:jfactormaps}, we create a flux map from our pixel-by-pixel calculation of the $\mathcal{J}$-factor. Since the simulated Sgr source is not smooth, we smooth the counts maps obtained from the simulation with a Gaussian kernel of size 0.2$^{\circ}$, ensuring that the total $\mathcal{J}$-factor for each map is preserved. We find that this smoothing step is necessary to obtain a good model of the source, which is generally best fit by a point source model, or a smooth model with a small, $\lesssim$ 1 degree extension~\cite{Evans:2022zno}. 
Since Sommerfeld sources are the most promising candidates for detection, we focus on these models for the remainder of this section. 
For this portion of the analysis, we focus on the results of the Sgr analogue with the smallest angular size, Sgr 25 (angular size $\sim 1.4^{\circ}$), and compute TS values for the Sommerfeld model.
Choosing the Sgr analogue with the smallest angular size ensures the best fit to the source given its small extension, as attempting to fit to the other larger analogues produces poor overall fit qualities within the ROI.
For the Sommerfeld model of Sgr 25, we find a TS $\sim 7$, which corresponds to only $\sim 2\sigma$ preference for the extended source model. This includes both bound and unbound DM particles, as described above.
For a map which includes only the bound DM particles to the Sgr analogue, we again find a TS of $\sim 7$.

\par Though as discussed above we do not include MW DM emission in our source maps for the results presented above, we do check to see the impact of including this component in our fits. Practically speaking, when attempting to include the MW halo emission in our models, we find either TS values for the Sgr source of $\sim 0$, or very large TS values which result from the source absorbing an unphysically large amount of diffuse emission over a wide area.

\section{Discussion and conclusions}
\label{sec:discussion}

\par Using analogues of the Sgr dSph identified in the Auriga simulations, we have computed $\mathcal{J}$-factors for general models in which the annihilation cross section is velocity-dependent. We provide predictions for both the DM component that is associated with the Sgr dSph, and with the MW foreground component along the line-of-sight to the dSph. 
For the Sommerfeld model, we find that emission from the central source is generally larger than that from the foreground MW component in the central region ($\lesssim 0.5^\circ$) and therefore would most likely be identifiable as a point source in the data. We further find that the  $\mathcal{J}$-factors that we compute from the simulations are in agreement with those obtained from equilibrium models presented in ref.~\cite{Evans:2003sc}, implying that the latter are accurate for the central core region of Sgr.

\par For p and d-wave models, for the first time we establish the existence of an enhancement factor due to unbound DM particles from the MW that are spatially overlapping with Sgr. These unbound particles have a relative velocity distribution characteristic of the smooth DM component of the MW at the radius of the Sgr analogue. The 
 enhancement factor results from annihilation of these unbound DM particles with themselves and with DM particles bound to the Sgr analogues. For p-wave models, we find that this 
enhancement  varies from $\sim 2-30$, depending on the exact Sgr analogue used. For d-wave models, we find a larger 
enhancement varying between 
$\sim 20-4000$. This 
enhancement is different than the more traditional ``boost" factor that has been discussed often in the literature, which results from DM substructure~\cite{Strigari:2006rd,Lacroix:2022cjm,Facchinetti:2022jtd}. 
On the other hand, for Sommerfeld models, the enhancement is  minimal since the unbound DM particles modify the relative velocity distribution at high speeds, which is suppressed due to the $1/v$ dependence of the Sommerfeld model. For the s-wave models, the enhancement is also minimal, due to the low density of the unbound DM particles compared to the bound DM particles.

\par Due to the angular extension of our models of $\sim 1-3$ degrees, we find that our Sgr templates generally do not provide as good a fit to the gamma-ray data as a simpler point source model. For example, examining in detail the Sommerfeld model for the case of our Sgr analogue with the smallest angular extent (Sgr 25, with an angular extent of 1.4$^{\circ}$), on Fermi-LAT data, we find a TS $\sim 7$ for the template including emission from bound and unbound particles. The TS remains the same when restricting to only the particles identified as bound to the Sgr analogue. We note that this TS value is less significant than the TS values for the point-like scenario for the M54/Sgr source as reported in refs.~\cite{20224FGL3DR}~and~\cite{Evans:2022zno}. More detailed modeling of combined emission from the MW halo and Sgr analogue DM annihilation would require specialized detailed diffuse model templates for the region, which is beyond the scope of this work.

An effect  not considered in this work is the presence of the Large Magellanic Cloud (LMC), which could have a substantial impact on the disruption of MW dwarf galaxies~\cite{Vasiliev:2021}. It is in general difficult to find exact LMC analogues in cosmological simulations. In particular, in the  Auriga simulations, LMC analogues are identified only at higher redshifts and within the last 8 Gyrs, but not at the present day~\cite{Smith-Orlik:2023kyl}. However, since in this work we match the properties of the Sgr analogue to its observed properties, we do not expect the existence of an LMC analogue in the simulations to significantly impact our results.

\par We finally note that it is likely that the  
enhancement factor that we derive is not restricted to just Sgr-like systems. In fact, it should exist for all subhalos that are within close enough proximity to the Galactic center where there is a significant DM density that is spatially coincident with the subhalo or for subhalos that are in an overdense region in the Galaxy. Furthermore, a large difference between the average relative velocity of the DM particles bound to the subhalo and to the smooth MW halo in that region also contributes to this 
enhancement factor. Particularly interesting systems to investigate in the future include Segue 1~\cite{Essig:2010em}, Omega Centauri~\cite{Evans:2021bsh}, or the newly-discovered Ursa Major III~\cite{Crnogorcevic:2023ijs}. The 
enhancement happens to be particularly prominent for Sgr-like systems due to both their large spatial extension and the proximity to the Galactic center. Quantifying this 
enhancement factor for subhalos of different size and at different distances will be a topic for future analysis. 
 
\acknowledgments
We thank Robert Grand for discussions on this paper. EV and NB acknowledge the support of the Natural Sciences and Engineering Research Council of Canada (NSERC), funding reference number RGPIN-2020-07138, the NSERC Discovery Launch Supplement, DGECR-2020-00231, and the HQP Pooled Resources from the McDonald Institute, HQP 2020-R6-01. AE, OH, and LS acknowledge support from DOE Grant de-sc0010813, NASA grant 80NSSC22K1577, and by the Texas A\&M University System National Laboratories Office and Los Alamos National Laboratory. NB acknowledges the support of the Canada Research Chairs Program. AHR is supported by a Research Fellowship from the Royal Commission for the Exhibition of 1851. NS is supported by an NSF Astronomy and Astrophysics Postdoctoral Fellowship under award AST-2303841. We have used simulations from the Auriga Project public data release~\citep{Grand:2024} available at https://wwwmpa.mpa-garching.mpg.de/auriga/data. This work was completed using the \textsc{Python} programming language as well as the following software packages: \textsc{astropy} ~\citep{astropy:2018}, \textsc{pandas}~\citep{reback2020pandas}, \textsc{numpy}~\citep{         harris2020array}, \textsc{scipy}~\citep{2020SciPy-NMeth}, \textsc{matplotlib}~\citep{Hunter:2007}, and \textsc{spyder}~\citep{raybaut2009spyder}.
\bibliographystyle{JHEP}
\bibliography{biblio}

\end{document}